\documentclass[12pt]{article}
\usepackage{caption}
\usepackage{graphicx}
\usepackage{amssymb}
\usepackage[round]{natbib}
\def\be{\begin{equation}}
\def\ee{\end{equation}}
\def\ba{\begin{eqnarray}}
\def\ea{\end{eqnarray}}

\usepackage{setspace}
\begin{document}
\begin{center}
{\bf Cloud base signature in transmission spectra of exoplanet atmospheres}\\

Sanaz Vahidinia$^1$, Jeffrey N. Cuzzi$^2$,  Mark Marley$^2$, Jonathan Fortney$^3$ \\
\end{center}
$^1$Corresponding author; Bay Area Environmental Research Institute, NASA Ames Research Center, Moffett Field, CA; sanaz.vahidinia@nasa.gov;
$^2$Space Science Division, Ames Research Center, Moffett Field, CA; 
$^3$Astronomy Dept., University of California, Santa Cruz CA 

\begin{center}
\vspace{2 mm}
\today
\end{center}

\vspace{10 mm}
\begin{abstract}
We present an analytical model for the transmission spectrum of a transiting exoplanet, showing that a cloud base can produce an observable inflection point in the spectrum. The wavelength and magnitude of the inflection can be used to break the degeneracy between the atmospheric pressure and the abundance of the main cloud material, however the abundance still depends on cloud particle size. An observed inflection also provides a specific point on the atmospheric P-T profile,  giving us a ``thermometer" to directly validate or rule out postulated cloud species. We apply the model to the transit spectrum of HD189733b.
\end{abstract}

\section{Introduction}
Transmission spectra of transiting exoplanets can provide a constraint on the position and type of cloud layers, temperature profiles, and ultimately, the abundance of cloud-forming species (Seager and Sasselov 2000;  Pont et al. 2007; Bean et al. 2010; Moses et al. 2013; Kreidberg et al. 2014).
Perhaps the most convincing evidence of high altitude hazes on any extrasolar planet to date is found in the case of the transiting hot Jupiter HD 189733b, which 
is a 1.1-Jupiter mass planet orbiting a bright nearby K star and an excellent target for detailed atmospheric studies. This planet is notable because its transit radius follows a smooth wavelength dependence (Pont et al. 2008, 2013; Sing et al 2011), and lacks signatures of molecular or atomic absorption, although Na and K are detected at high spectral resolution (Pont et al. 2013, Huitson et al. 2012). A currently popular explanation for the HD 189733b transmission spectrum is a haze of small particles, extending five atmospheric scale heights (Pont et al. 2013), that are better scatterers than absorbers (Lecavelier des Etangs et al. 2008, henceforth L08); that is, the particle's  scattering efficiency is large compared to its absorption efficiency. This property, providing a $\lambda^{-4}$ opacity dependence, requires a material with an extremely small imaginary index of refraction (to avoid dominance by absorption, with its $1/\lambda$ wavelength dependence;  see section 2) and $\rm MgSiO_{3}$ was suggested as a candidate. Other silicates, however, also condense in similar regions, as does Fe, so further exploration of this hypothesis seems appropriate. 
Here, we present a model for interpreting the transmission spectra of transiting planets, extending the models of Fortney (2005; henceforth F05) and L08 to incorporate condensation cloud bases. 

\section{Planet and cloud geometry}

We consider an atmosphere containing a cloud formed from condensed particles at an altitude defined by the intersection of the T-P profile of the atmosphere and the  condensation vapor pressure of the material. The typical vertical structure of a cloud is a sharp base at the condensation temperature and pressure (T,P), with a more gradual vertical falloff in density (Ackerman and Marley 2001, henceforth AM01).  For simplicity we have assumed a single condensate.

\begin{figure*}[!h]
\centering
\includegraphics[angle=0,width=4.0in,height=2in]{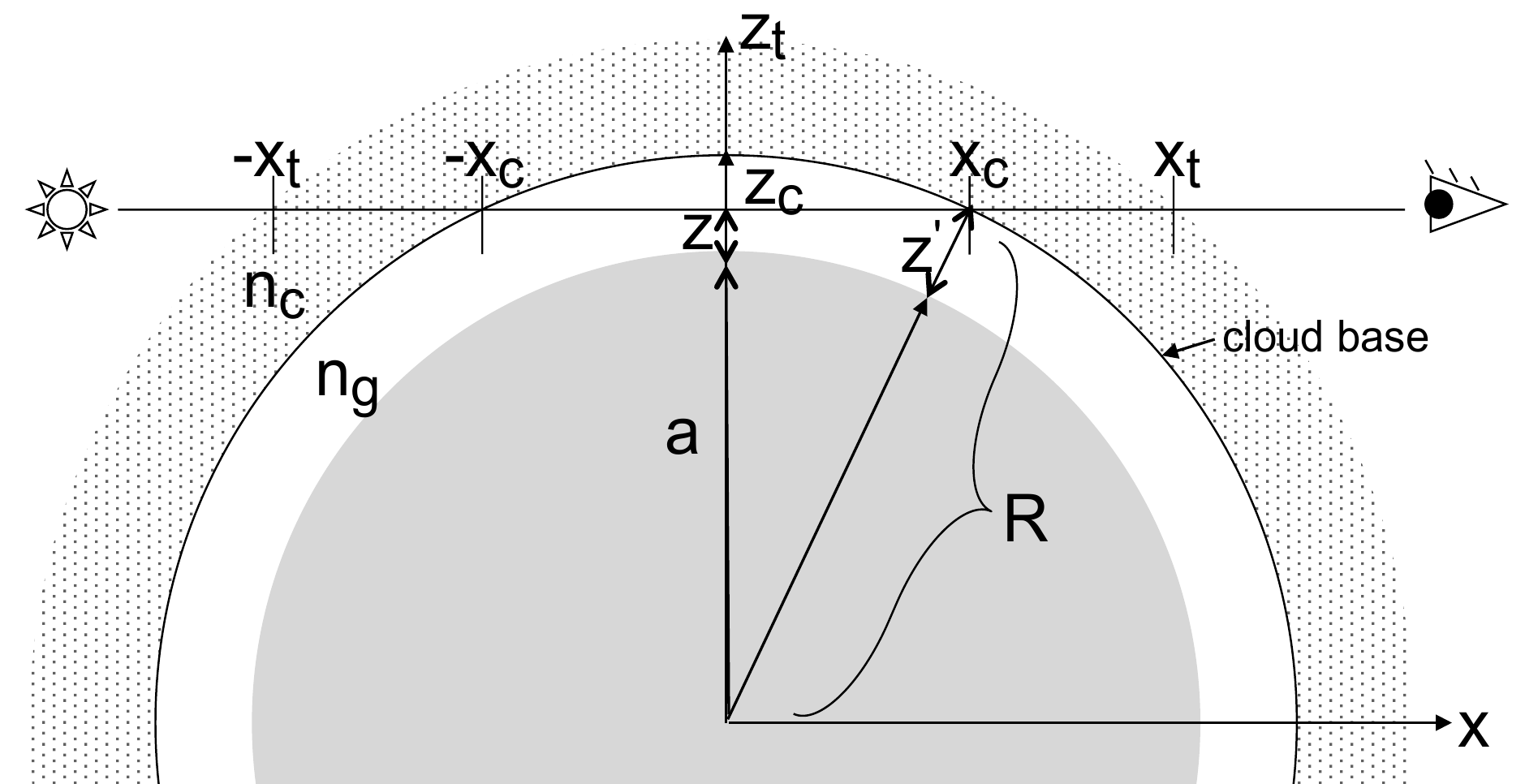}
\caption{Atmosphere with cloud base:  The gas constituent extends throughout the atmosphere and the particle cloud or haze (stippled) has base at $z_c$ (black curve) and top at $z_t$.  The radial distance to any point in the atmosphere is $R=a+z^{'}$; a line of sight is characterized by its minimum altitude $z$, and a horizontal distance along this path is given by $x =\sqrt{2a(z^{'}-z)}$; for $z\ll a$. If the cloud evaporates at a higher altitude, perhaps due to a temperature inversion, then we can also define a cloud top $z_t$, with $x_t=\sqrt{2a(z_t-z)}$. }
\label{cartoon}
\end{figure*}

A photon that traverses this atmosphere along the line of sight in transmission geometry  is affected by gas and cloud particle opacities. However, the opacity for photons with  wavelengths that can reach altitudes underneath the cloud base is reduced as it crosses the region below the cloud base, where only gas opacity contributes. This gap in opacity between $-x_{c}$ and $x_{c}$ (see Figure \ref{cartoon}) shows up as an inflection point in the transmission spectrum, as described below.

The number density $n_{g}$ of gas molecules with scale height $H$ is a function of radial distance $R$ from the center of the planet, as measured relative to some reference radius $a$:
\begin{equation}
 {n_{g}(R) = n_{0}e^{-(R-a)/H}} = n_{0}e^{-z^{'}/H}
\end{equation}
The cloud particle number density $n_{c}$ relative to $\rm H_{2}$ is defined by a number abundance $\zeta$, and the particles are assumed to have the gas scale height because small particles such as observed (L08) are easily lofted by atmospheric turbulence (AM01):
 \begin{equation}
n_{c}(R) =  \zeta n_g(R) = \zeta n_{0} e^{-(R-a)/H}
\end{equation}
For the case of $\rm MgSiO_{3}$ (enstatite), we follow L08 in assuming it is limited by the number of magnesium atoms; thus 
\begin{equation}\label{eq:zetamg}
\zeta=2\zeta_{Mg} \mu_{\rm MgSiO_{3}} m_{p}/ \mu_{\rm H} \rho V,
\end{equation}
where $\zeta_{\rm Mg}$ is the number abundance of Mg relative to $\rm H_{2}$ in the planets atmosphere, and its cosmic/solar value is $\zeta_{\rm Mg}^{*} \sim 4 \times 10^{-5}$. The enstatite molar mass $\mu_{\rm MgSiO_{3}}$ is $100.4$ times the hydrogen molar mass $\mu_{\rm H}$,  $m_{p}$ is the mass of the proton, the enstatite grain density is $\rho =3.2 \rm g \, cm^{-3}$, and the particle volume is $V=4\pi r^{3}/3$.

Following F05, we can calculate the horizontal optical depth $\tau_{h}(z)$ along a line of sight (in the $x$ direction) for an atmosphere with one cloud base (Figure \ref{cartoon}), as the sum of contributions from gas and cloud particles ($j$). 

\begin{equation}
\tau_{h}(z, \lambda) = \sum_{j}  \int_{-\infty}^\infty {n_{j}(x) \sigma_{j}(\lambda) dx}  = \int_{-\infty}^{\infty}{n_{g}(x) \sigma_{g}(\lambda) dx} +
\int_{-\infty}^{\infty} {n_{c}(x) \sigma_{c}(\lambda) dx}, \mbox{    if } z>z_{c},
\end{equation}

\begin{equation}
\tau_{h}(z, \lambda) = \sum_{j}  \int_{-\infty}^\infty {n_{j}(x) \sigma_{j}(\lambda) dx}  = \int_{-\infty}^{\infty}{n_{g}(x) \sigma_{g}(\lambda) dx} +
2\int_{x_{c}}^{\infty} {n_{c}(x) \sigma_{c}(\lambda) dx}, \mbox{    if } z<z_{c},
\end{equation}
where $n_{j}(x)=n_{j}(z) e^{-x^{2}/2aH}$, using $R=z^{'}+a$ and $z^{'}=(x^{2} + 2az)/2a$ as shown in Figure \ref{cartoon}.
The extinction cross section $\sigma(\lambda)$ is defined as $\sigma(\lambda)=Q_{e} \, \pi r^{2}$ and $Q_{e}(\lambda)=Q_{a}(\lambda) + Q_{s}(\lambda)$, where  
the absorption ($Q_{a}$) and scattering ($Q_{s}$) 
efficiencies are functions of both particle radius $r$ and wavelength $\lambda$, but we suppress the $r$-dependence notation. Following F05, $\tau_{h}(z)$ can then be written as:

\begin{equation}\label{eq:tauh5}
\tau_{h}=  \sigma_{g}(\lambda) n_{0} e^{-{z \over H}} \sqrt{2\pi a H} +  \sigma_{c}(\lambda) \zeta n_{0} e^{-{z\over H}} \sqrt{2\pi a H}, \mbox{    if } z>z_{c}
\end{equation}

\begin{equation}\label{eq:tauh6}
\tau_{h}=  \sigma_{g}(\lambda) n_{0} e^{-{z \over H}} \sqrt{2\pi a H} +  \sigma_{c}(\lambda) \zeta n_{0} e^{-{z\over H}} \sqrt{2\pi a H}  \left(1-{\rm erf}\left( {x_{c} \over \sqrt{2aH}} \right)\right), \mbox{    if } z<z_{c},
\end{equation}
where from geometry, $x_{c}=\sqrt{2a(z_{c}-z)}$ (Figure \ref{cartoon}, assuming $(z,z^{'})\ll a$). L08 calculated a nearly model-independent equivalent optical depth $\tau_{eq} \approx 1/2$ for a range of atmospheric scale heights, where the planet with its translucent atmosphere produces the same absorption depth as a sharp occulting disk. This 
approximation is valid either above or below a cloud base. We can rewrite equation \ref{eq:tauh5} using the ideal gas law to express the pressure $P(z)=P_{0}e^{-z/H}$ of the hazy atmosphere above the cloud base and at $\tau=\tau_{eq}$ as:
\begin{equation}\label{eq:oldeq8}
P(z,\zeta,T,\sigma_{c})={k T \tau_{eq} \over \sqrt{2\pi a H}(\sigma_{g}(\lambda)+\zeta \sigma_{c}(\lambda))}
\end{equation}

\noindent The gas reference pressure $P_{0}$ is chosen at reference wavelength $\lambda_{0}=750$nm ($\lambda_{0}$ is arbitrary and we choose this value to be consistent with L08) where the altitude $z=0$ or $R=a$, and $P_{0}=kT\tau_{eq}/\sigma_g(\lambda_{0}) \sqrt{2\pi a H}$.

The altitude $z$ at which $\tau_h(z,\lambda)=\tau_{eq}$ is then (from equation \ref{eq:tauh6}):
\begin{equation}\label{eq:hlnz}
z = H \mbox{ } \ln \left[{P_{0} \over k T \tau_{eq}} \sigma_{g}(\lambda)  \sqrt{2\pi a H} +   {\zeta P_{0} \over k T \tau_{eq}} \sigma_{c}(\lambda)  \sqrt{2\pi a H}  \left(1- {\rm erf}\left({x_{c}\over \sqrt{2aH}} \right)\right)\right]
\end{equation}

For a single component atmosphere where either gas or particles dominate, equation \ref{eq:hlnz} can be written (neglecting constants) as $ z \sim H \ln (\lambda^{-\alpha}) \sim -H \alpha \ln(\lambda)$, 
where $\sigma(\lambda) \sim \lambda^{-\alpha}$. Therefore for a constant scale height, the transit curve is a log-$\lambda$ function with slope 
ranging from flat for large particles ($\alpha=0$), to a steeper value for gas molecules or non-absorbing Rayleigh scatterers with constant real refractive index ($\alpha=4$). The scale height $H$ also contributes to the slope (L08).  

To the extent that the particle refractive indices are wavelength dependent,  more detailed variations can occur in the transit spectrum.
For instance, small particles in the Rayleigh regime show different wavelength and particle size dependence if they are essentially lossless ({\it eg. } van de Hulst 1957): 
\begin{equation}\label{eq:sigmasubs}
\sigma_c(r,\lambda) \approx Q_s \pi r^2 = {24 \pi^3 (n_r^2-1)^2 \over \lambda^4 (n_r^2+2)^2 }V^2
\end{equation}
or if they are even slightly absorbing:
\begin{equation}
\sigma_c(r,\lambda) \approx Q_a \pi r^2 = { 36 \pi n_r n_i \over \lambda (n_r^2+2)^2} V.
\end{equation}
Where $n_{r}$, and $n_{i}$ are the real and imaginary refractive indices. The extinction cross section for $\rm H_{2}$ gas is (L08):
\begin{equation}
\sigma_g(\lambda)  = \sigma_{0}\left({\lambda \over \lambda_{0}}\right)^{-4},\mbox{where $\sigma_{0}=2.52\times10^{-28}$cm$^{2}$ and $\lambda_{0}=750$nm} 
\end{equation}

\section{Model Results}

\subsection{Inflection points}

\begin{figure*}[h!]
\begin{center}
\includegraphics[angle=0,width=3.2in,height=2.3in]{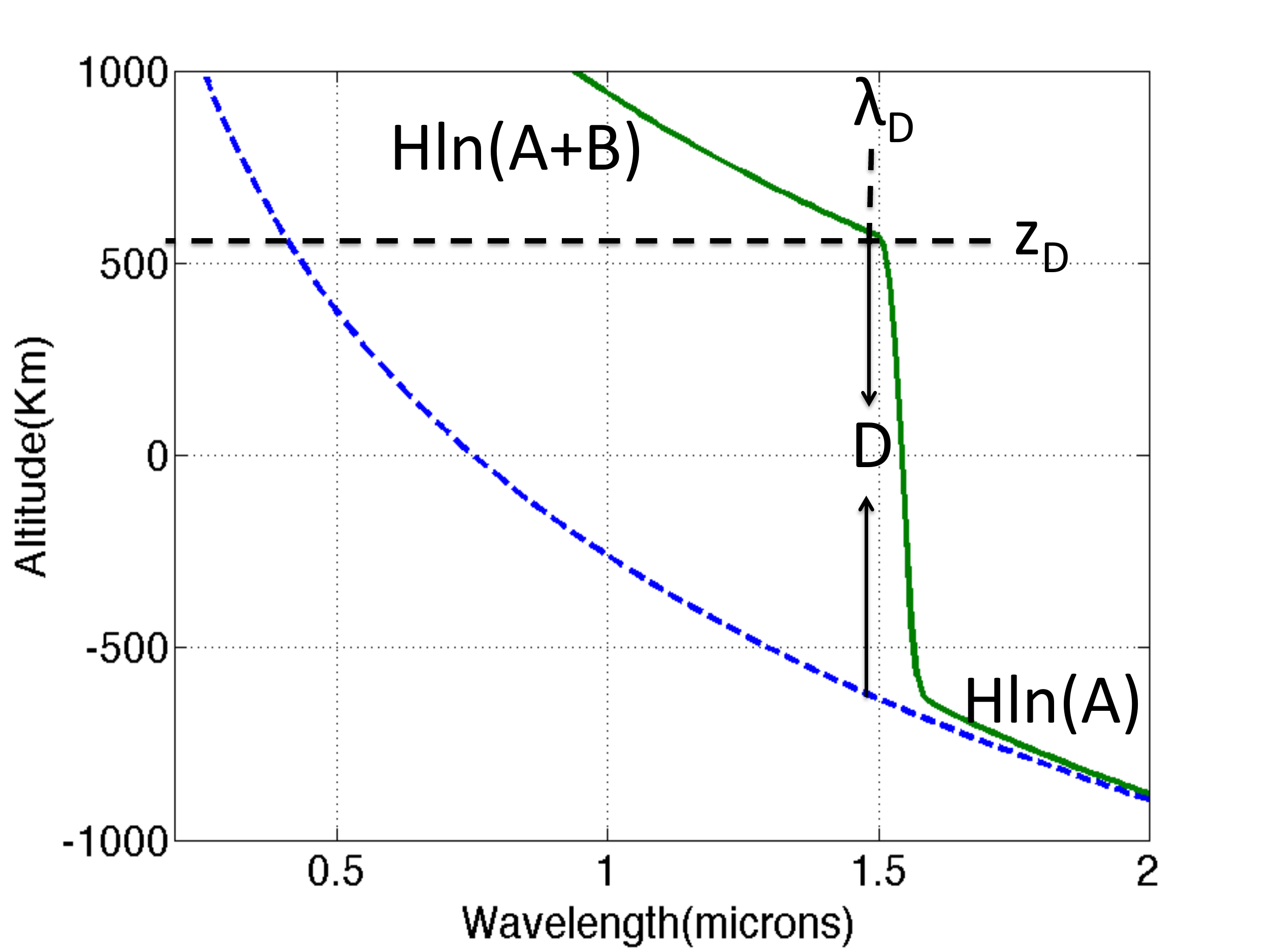}
\includegraphics[angle=0,width=3.2in,height=2.3in]{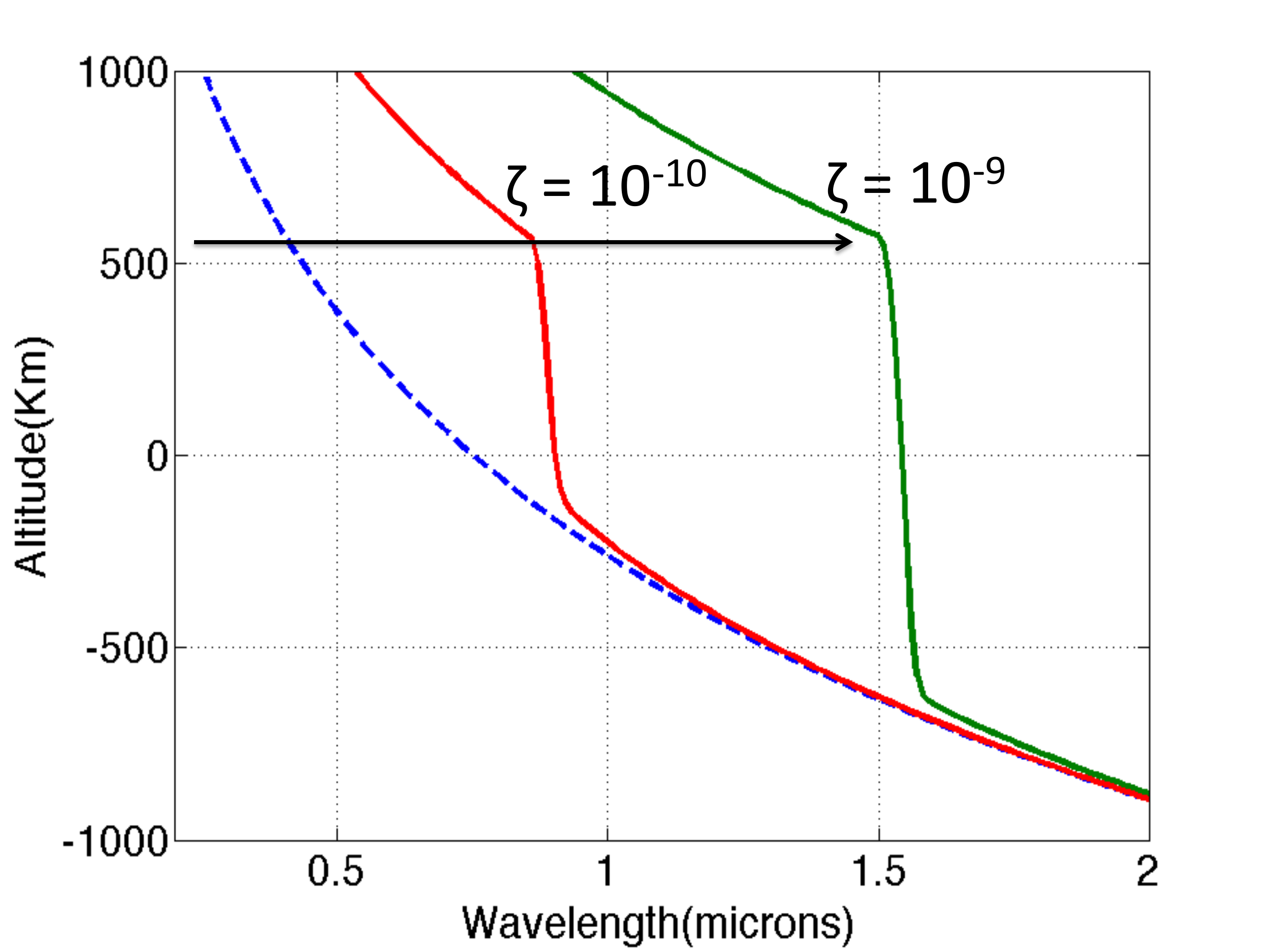}

\caption{(left) The transit altitude profile showing transition from the spectral region where cloud particle opacity dominates ((green) $z = H \ln (A+B)$) to the region sampling beneath the cloud base, where gas opacity dominates ((blue) $z = H \ln (A)$). The inflection point defines two critical parameters -  the wavelength $\lambda_{D}$ and the drop $D$ (see section 3). The value of $\lambda_D$ depends on the pressure at the cloud base. (Right) The value of $\lambda_D$ also depends on particle abundance $\zeta$, and a higher $\zeta$ also results in a larger drop $D$.}
\label{inflection}
\end{center}
\end{figure*}

We illustrate the observable effect of a cloud base with a simple toy model calculation using equation \ref{eq:hlnz}, with a constant temperature atmosphere consisting of Rayleigh scattering H$_{2}$ molecules, and $\rm MgSiO_{3}$ grains for the cloud constituent. In Figure \ref{inflection} (left), we show the altitude (related to apparent radius) at which $\tau_{h} = \tau_{eq}$. Results are shown for a cloud base at an arbitrary altitude, for constant $H$, and two abundances corresponding to the inflection points at two different wavelengths $\lambda_D$ (Figure \ref{inflection} (right)). For lines of sight passing above the cloud base (altitudes $z>z_{c}$, sensed by wavelengths $< \lambda_{D}$), both gas and cloud particles contribute to the opacity, leading to more extinction at shorter wavelengths than the gas-only curve (dotted blue). On lines of sight with altitudes below the cloud base, sensed by wavelengths $> \lambda_{D}$, primarily gas opacity contributes and the red curve falls onto the gas curve.  In section \ref{TPfits} we will show the effect of realistic T-P profiles.

A cloud base inflection point, if observed, has interesting implications about cloud particle properties and atmospheric properties in general. To illustrate this, we rewrite equation \ref{eq:hlnz} in simpler form:
\begin{equation}
z = H \mbox{ } \ln \left[  A +   B\left(1-{\rm erf}\left({x_{c}\over \sqrt{2aH}} \right)\right)\right]
\end{equation}

\begin{equation}\label{eq:oldeq12}
A= {\sigma_{g}(\lambda) P_{0} \sqrt{2\pi a H} \over k T \tau_{eq}}  , \mbox{ }
B= {\sigma_{c}(\lambda) \zeta P_{0} \sqrt{2\pi a H}  \over k T \tau_{eq}}   , \mbox{ }
\end{equation}

\begin{equation}\label{eq:oldeq14}
{B \over A}= {\sigma_{c}(\lambda) \zeta \over \sigma_{g}(\lambda)}, \mbox{ and } D=H \ln \left[1+{B\over A}\right] 
\end{equation}

\noindent where $D$ is the height of the inflection step. In Figure \ref{inflection} (left panel) the transit altitude profile is broken up into two regions: the spectral region where gas and cloud particle opacities contribute (green curve at $\lambda < \lambda_D$) is approximated by $H \ln (A+B)$  and the region sampling beneath the cloud base, where gas opacity dominates, by $H \ln (A)$. The drop at $\lambda_D$ from $H \ln (A+B)$ to $H \ln (A)$, corresponding to detection of the cloud base, is the inflection height $D$. Shown in Figure \ref{inflection} (right panel) is the shift of the inflection point in wavelength depending on particle abundance $\zeta$ while keeping the cloud base pressure constant. Higher $\zeta$ results in larger opacity, so longer $\lambda$ is needed to reach the altitude/pressure where the cloud base is located; this then also results in a larger value of $D$ (equation \ref{eq:oldeq14}). 

The two critical properties - the drop $D$ and the wavelength $\lambda_{D}$ - allow us to determine the cloud material abundance (a function of particle size) and cloud base pressure. The abundance $\zeta$, $P_{0}$, and $\sigma_{c}$ are degenerate in the term $B$ above, but if we can detect a cloud base in the transit spectrum, have a candidate condensate and T-P curve, and assume a particle radius $r$ to calculate $\sigma_{c}$, then the degeneracy can be broken and limits can be placed on the abundance $\zeta$ for the candidate material as shown in section 4. 

\subsection{Cloud base and cloud top}
\begin{figure*}[!h]
\begin{center}
\includegraphics[angle=0,width=4in,height=2.6in]{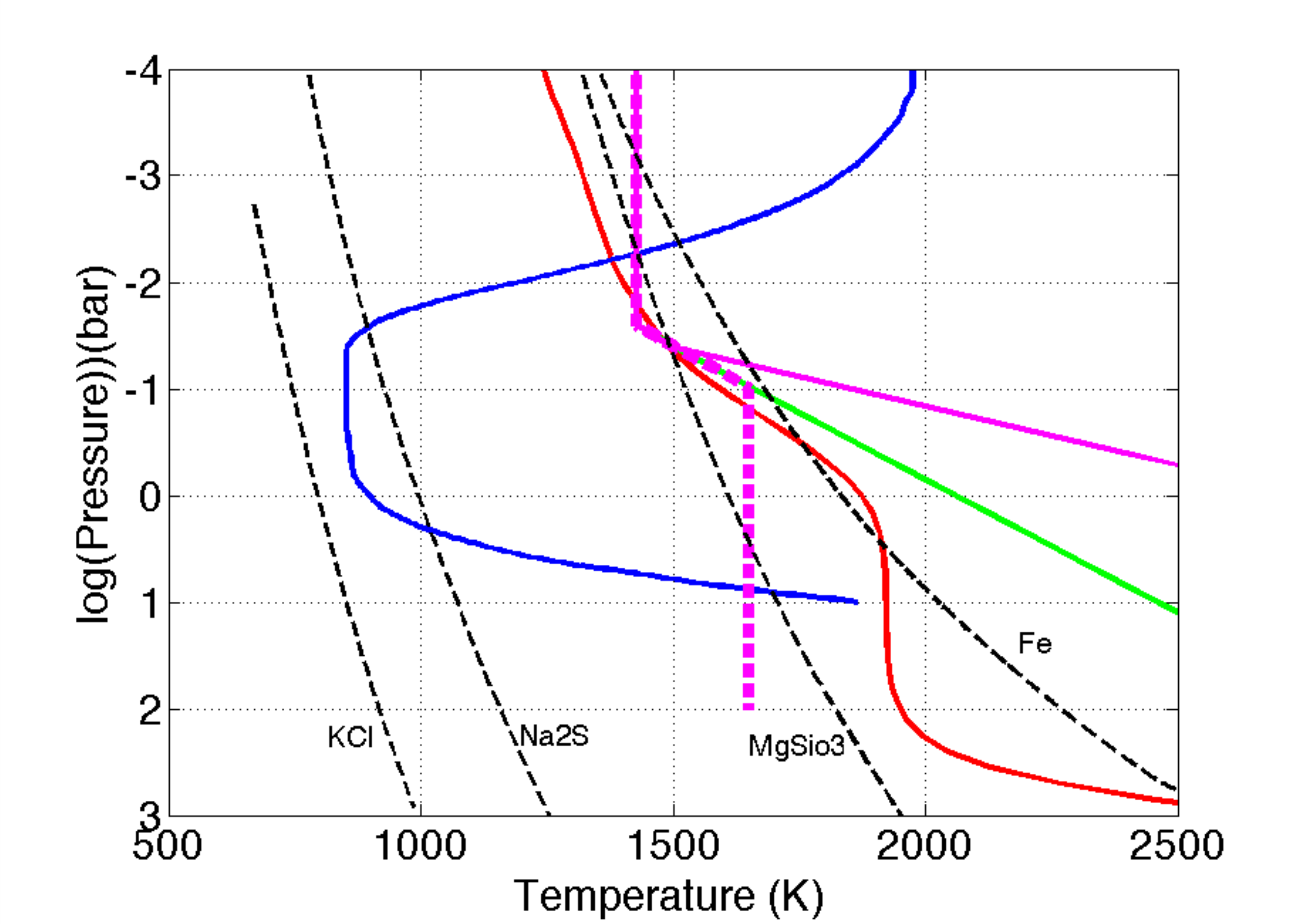}
\caption{\small{Alternate T-P profiles for HD189733b: the red curve is shifted by $+250$K from Fortney et al. (2008; see section \ref{sec:Data}) and the blue curve is from Pont et al. (2013). The green and magenta curves are artificial profiles we constructed to study the effect of the temperature gradient on the spectrum. We also show condensation curves (dashed black) for potential cloud forming candidate materials; cloud bases form when T-P curves intersect these vapor pressure curves.}}
\label{TP}
\end{center}
\end{figure*}

Figure \ref{TP} shows two candidate T-P profiles for HD189733b and several arbitrary profiles. Even the two actual candidates are very different, indicative of the current uncertainties.  We also show vapor pressure curves for several different materials; a cloud base forms when the decreasing temperature with altitude crosses a vapor pressure curve. Moreover, we could have a cloud with well-defined base {\it and} top, as in the case of the $\rm MgSiO_{3}$ condensation line crossing the blue T-P profile with inversion,    at two places (Figure \ref{TP}, T-P profile is from Pont et al. (2013) using a model from Heng et al. (2012)).  For simplicity we can break up the calculation for this scenario into three opacity regions: as wavelength increases,  photons traverse deeper into the atmosphere. They first encounter purely gas opacity, then below the cloud top gas and condensate opacities both contribute, and finally below the cloud base gas opacity dominates again.

Region 1, $z>z_{t}$:
\begin{equation}
z \sim H \mbox{ } \ln\left[{P_{0} \over k T \tau_{eq}} \sigma_{g}(\lambda)  \sqrt{2\pi a H}\right]
\end{equation} 

Region 2, $z_{c}<z<z_{t}$:
\begin{equation}
z \sim H \mbox{ } \ln\left[{P_{0} \over k T \tau_{eq}} \sigma_{g}(\lambda)  \sqrt{2\pi a H} +   {\zeta P_{0} \over k T \tau_{eq}} \sigma_{c}(\lambda)  \sqrt{2\pi a H}  \left({\rm erf}\left({x_{t}\over \sqrt{2aH}} \right)\right)\right]
\end{equation} 

Region 3, $z<z_{c}$:
\begin{equation}\label{eq:oldeq11}
 z \sim H \mbox{ } \ln\left[{P_{0} \over k T \tau_{eq}} \sigma_{g}(\lambda)  \sqrt{2\pi a H} +   {\zeta P_{0} \over k T \tau_{eq}} \sigma_{c}(\lambda)  \sqrt{2\pi a H}  \left({\rm erf}\left({x_{t}\over \sqrt{2aH}} \right)-{\rm erf}\left({x_{c}\over \sqrt{2aH}} \right)\right)\right]
\end{equation}





\section{Model comparisons with data}
\label{sec:Data}
\begin{figure*}[h!]
\centering
\includegraphics[angle=0,width=4.0in,height=3in]{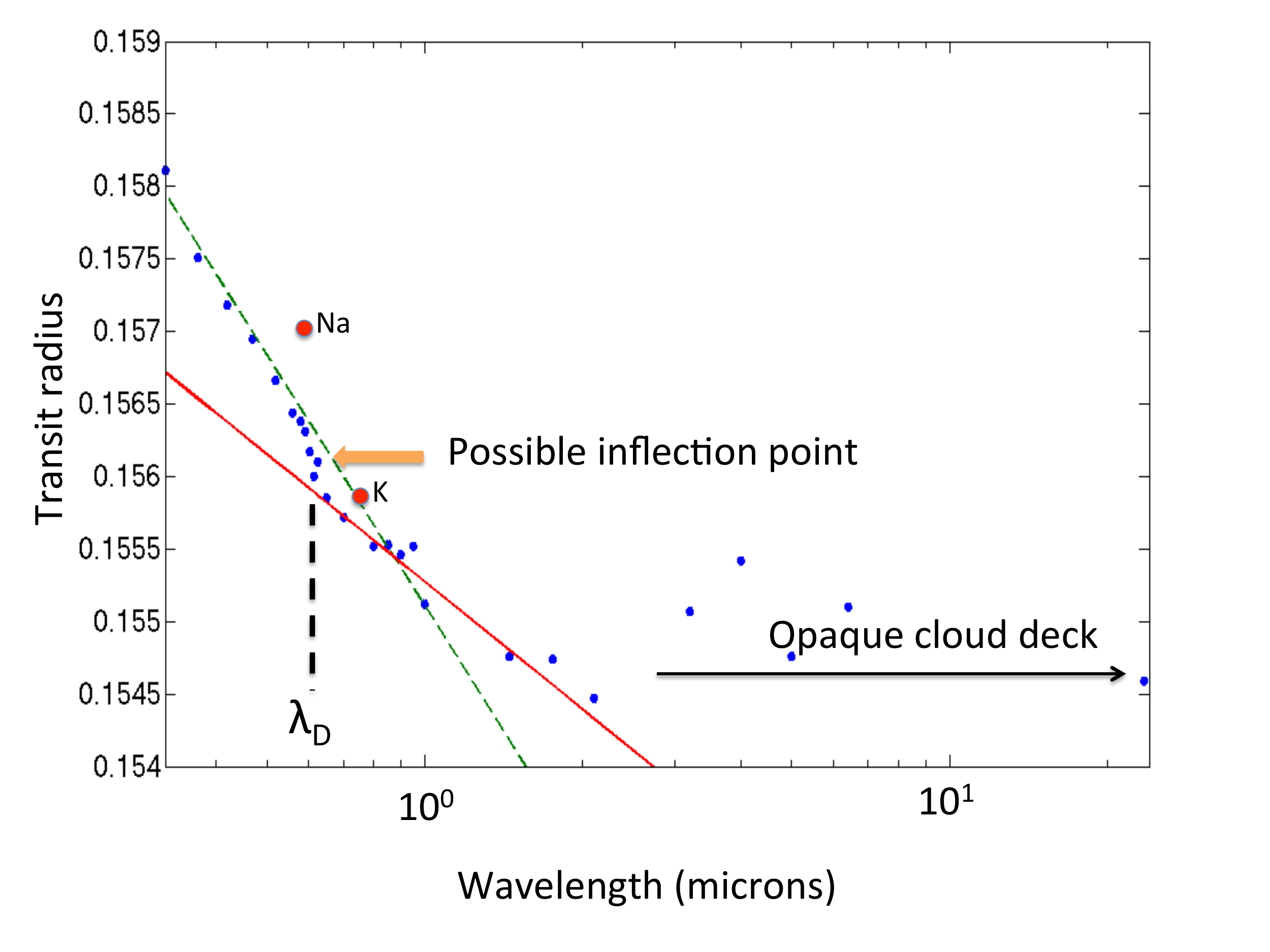}
\caption{HD189733b data (taken directly from table 6 of Pont et al. 2013, black) compared with arbitrary log functions (red and green) with two different slopes ($x$-axis is shown with log scale).
Na and K gas absorption lines are shown as red dots. Some have suggested that the data at $\lambda > 1 \mu$m suggest an opaque cloud deck. 
 }
\label{pontdata}
\end{figure*}

Recently, Pont et al. (2013) published a combined transit spectrum of HD189733b from their newly tabulated data from HST, ranging from UV to Infrared.  This data set covers a larger range of wavelengths, at better resolution than previous observations, and may show hints of inflection points. The spectrum can be described with a combination of two 
different slope $\ln(\lambda)$ curves as shown in Figure \ref{pontdata}, where the slope change near $0.6 \mu$m may also be marked by an apparent drop similar to our 
inflection height $D$. However, the temperature needed to match the slope at short wavelengths is 2000K (Pont et al 2013), too hot for $\rm MgSiO_{3}$ to condense. We return to this point below. 

\subsection{Summary of procedure}
Before comparing the observations with model transmission curves we summarize the procedure as follows:

1) Calculate cloud base temperature $T_{c}$ from the transit slope ($T_{c}$ needs to be consistent with the presumed cloud material).
\begin{equation}\label{eq:dzdlambda}
T_{c}={\mu g \over k} {dz \over d\lambda} \left({\mathrm{d}\ln(\sigma_{g} +\zeta \sigma_{c})\over \mathrm {d}\lambda}\right)^{-1}.
\end{equation}

2) Using equation \ref{eq:oldeq14} with the observed inflection point wavelength $\lambda_D$ and drop $D$, calculate $\zeta \sigma_{c}(\lambda_D)$. Since we don't know the size of the cloud particles (beyond the hypothesis that $r \ll \lambda$), the particle abundance and 
particle cross section remain degenerate:
\begin{equation}
\zeta \sigma_{c}=\sigma_{g}(\lambda_D)(\mathrm{e}^{D/H}-1).
\end{equation}

3) We then rewrite equation \ref{eq:zetamg}, using the product $\zeta \sigma_{c}$ from above, and substitute equation \ref{eq:sigmasubs}  (the nominal case where the particles are Rayleigh {\it scatterers}), to determine the chemical abundance of magnesium $\zeta_{Mg}$ in the cloud: 
\begin{equation}\label{eq:zetacalc}
\zeta_{Mg}={\rho \, \zeta \sigma_{c} \over 200 m_{p}} \left({V \over \sigma_{c}}\right)= {1{\rm cm} \over 64 \pi^4}  { (n_r^2+2)^2 \over (n_r^2-1)^2 }\left({\rho \zeta \sigma_c \over m_p}\right)\left({ \lambda_D \over 1\mu{\rm m}}\right)^4  \left({0.01\mu{\rm m} \over r}\right)^3
\end{equation}

4) Substituting $\lambda_D$, $T_{c}$ and $\zeta \sigma_{c}(\lambda_D)$ into equation \ref{eq:oldeq8}, with $\tau_{eq}$=0.5, we calculate the cloud base pressure $P_{c}$. Then we can calculate a value for the reference radius $a$ which we can use to fit our model to data at other wavelengths (given as apparent radius).  Specifically, $a = R_{\lambda_D} -z_{c}$ where $z_{c}=H\ln(P_0/P_c)$.

Using the observed slope for $\lambda < \lambda_{D}$, equation \ref{eq:dzdlambda} gives $T_{c}=2000$K, 
as in Pont et al (2013). Since this temperature is too hot to sustain an 
$\rm MgSiO_{3}$ cloud, we adopt a lower temperature $T_{c}=1400$K (consistent with the slope below the cloud base)
where $\rm MgSiO_{3}$ remains solid. This discrepancy might be explained by a combination of uncertainty in the observed slope or in the value of $g$, and neglect of higher order terms in equation \ref{eq:hlnz} propagating into equation \ref{eq:dzdlambda}.  The corresponding cloud base pressure from equation \ref{eq:oldeq8} is $P_{c}=10^{-1.3}$ bar. The magnesium abundance $\zeta_{Mg}$ calculated from equation \ref{eq:zetacalc} is $2-5$ orders of magnitude lower than cosmic abundance for haze particles with $r\sim 0.01-0.1 \mu$m. There will be variability in our calculated values due to observational noise, and error analysis is appropriate for future work.

{\subsection{Specific comparisons}
\label{TPfits}
Finally, we used various T-P profiles shown in Figure \ref{TP} to create cloud models for comparison with the data of Pont et al (2013), as shown in Figure \ref{datafit}. The first calculation in Figure \ref{datafit} (left) uses the red T-P profile from Figure \ref{TP}, which was shifted by 250K from Fortney et al (2008) so that the cloud base pressure, set by the altitude where the $\rm MgSiO_{3}$ condensation line crosses the T-P profile, is equal to our calculated $P_{c}$. We can see that the model inflection point coincides with the possible inflection point in the data, but the model slope at shorter wavelengths is too flat. We even used actual, wavelength-dependent refractive indices of $\rm MgSiO_{3}$ to get the fit shown, but these didn't change the slope much from some typical constant value. Moreover, the model slope at $1-2 \mu$m wavelengths is too {\it steep}. 

We created artificial temperature profiles to investigate the effect further. These artificial T-P profiles (Figure \ref{datafit}, center) also have their cloud base and top defined by the altitude/pressure where they cross the $\rm MgSiO_{3}$ condensation line.  The dotted magenta curve is the closest in spirit to the modified Fortney et al. (2008) T-P curve, but being colder, has a flatter slope at long wavelengths. However it is actually slightly cooler just above the cloud base, so fits less well at short wavelengths. 

\begin{figure*}[h!]
\begin{center}
\includegraphics[angle=0,width=2.1in,height=2.3in]{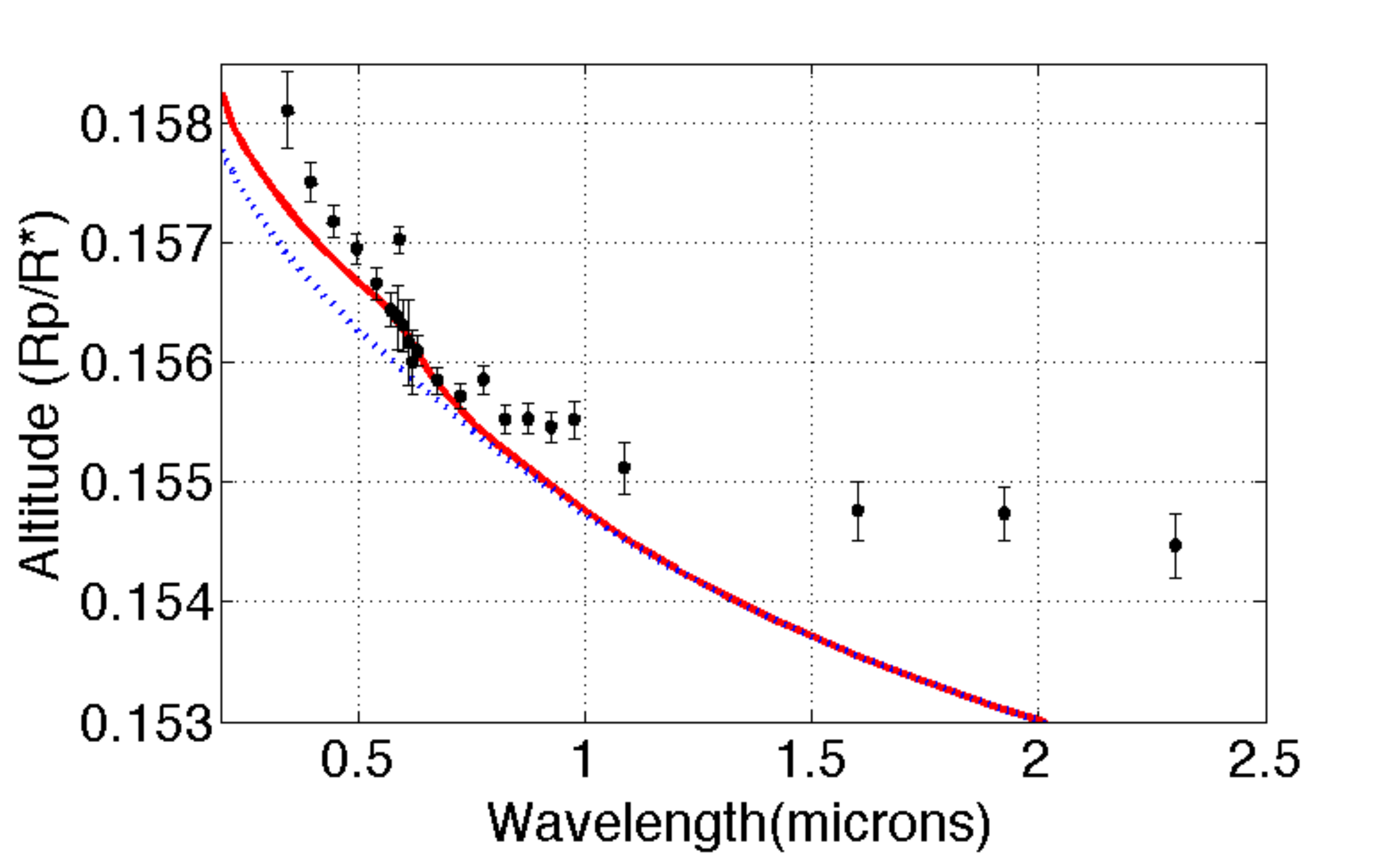}
\includegraphics[angle=0,width=2.1in,height=2.3in]{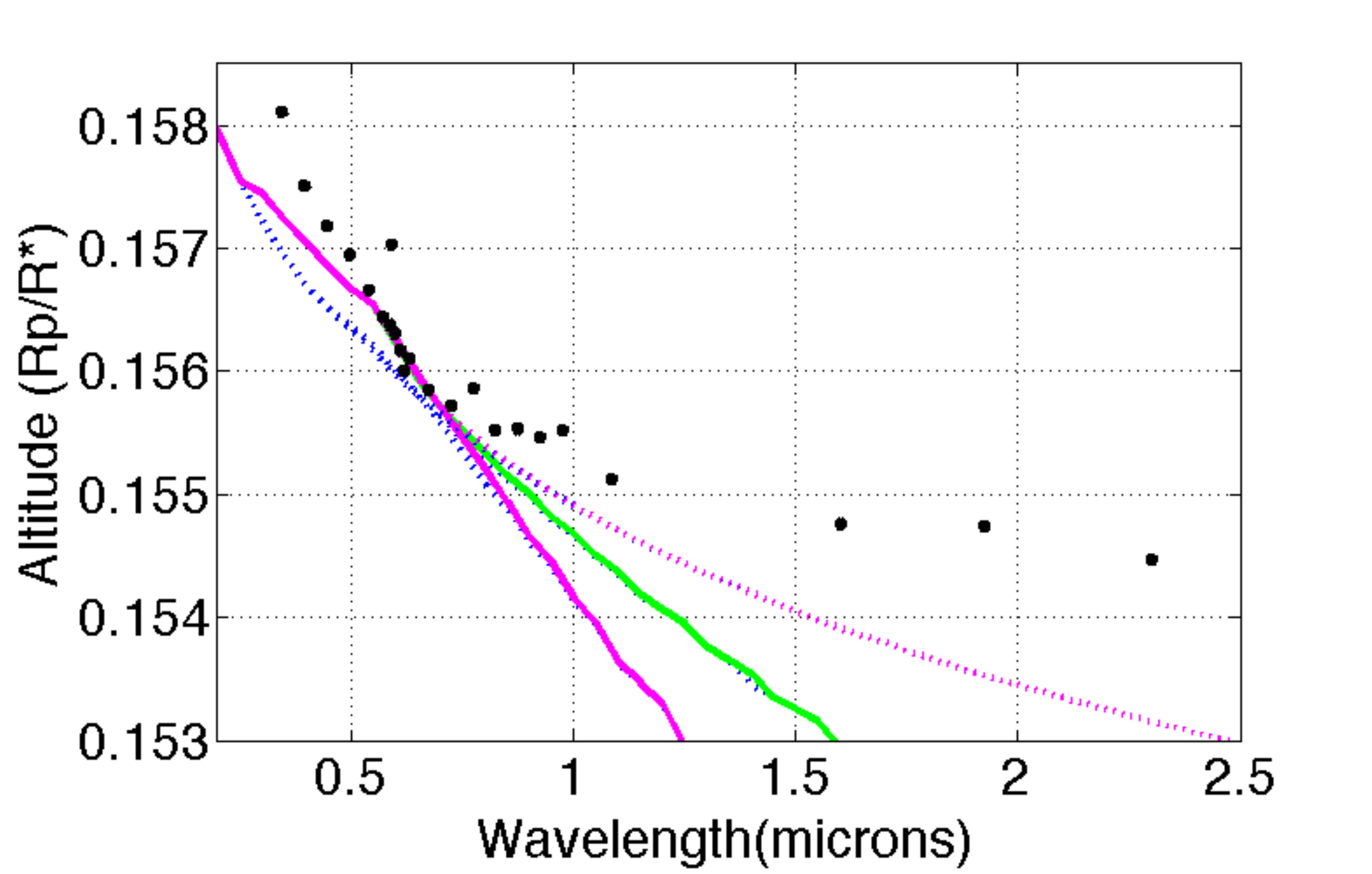}
\includegraphics[angle=0,width=2.1in,height=2.3in]{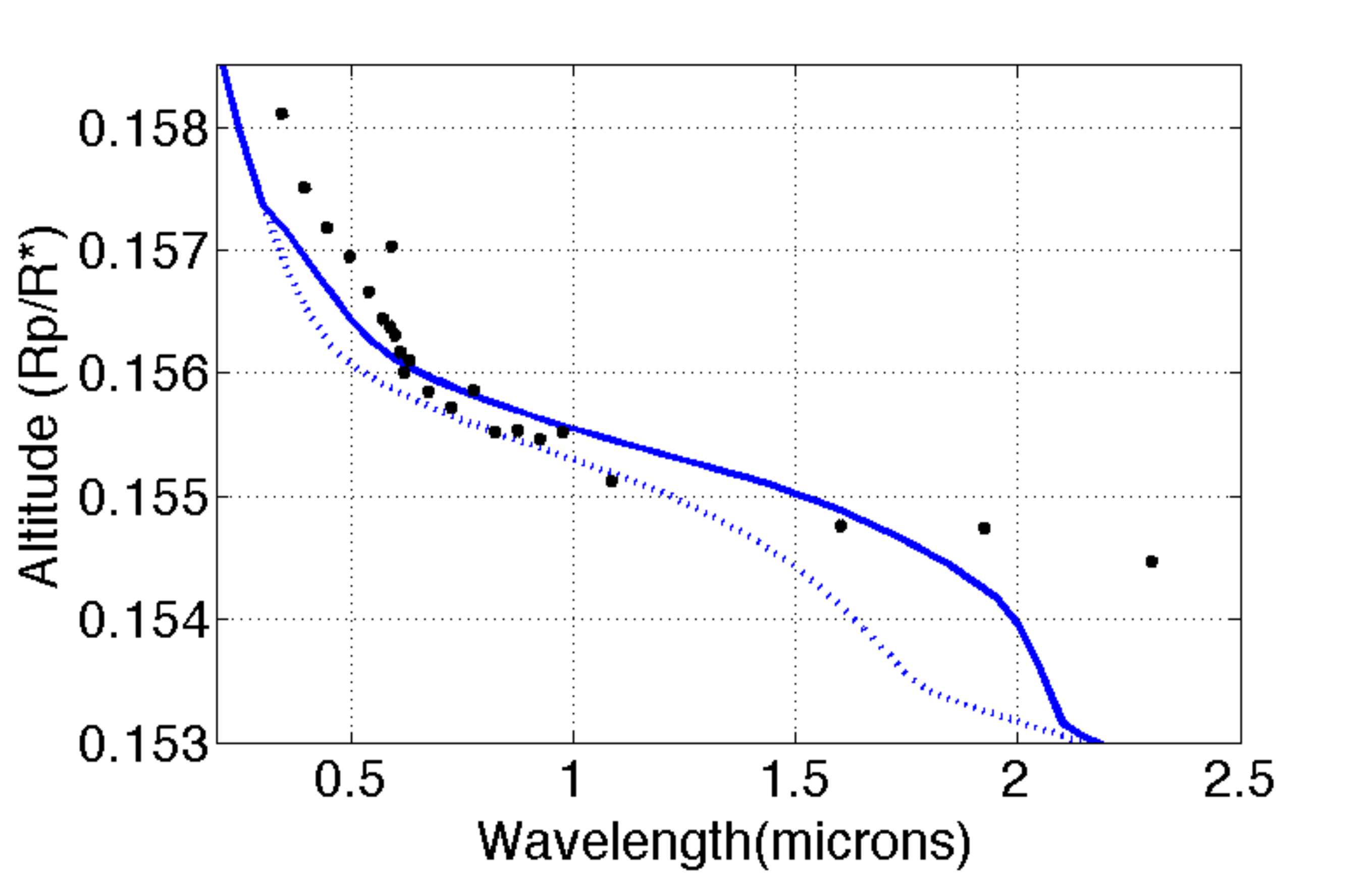}
\caption{(left) Ratio of planet radius $R_{p}$ to host star radius $R^*$ {\it vs.} wavelength. Data with error bars from Pont et al (2013) (black symbols) are fitted with our cloud model calculation using the red T-P profile in Figure \ref{TP} and a cloud base pressure of $\sim10^{-1.3}$ bar set by the altitude where the $\rm MgSiO_{3}$ condensation line crosses the red T-P profile. The gas curve (dotted blue) is also plotted for reference.  (Center) Pont et al. data is fitted with our cloud model calculation using the artificial profiles in Figure \ref{TP} with a cloud base and top pressures of $10^{-1.4}$ and $10^{-2.3}$ bar defined by the position where the $\rm MgSiO_{3}$ condensation line crosses the T-P profiles.  (Right) Pont et al. data is fitted with our cloud model calculation using the blue T-P profile in Figure \ref{TP} with cloud base and top pressures of $10^{0.8}$ and $10^{-2.3}$ bar set by  where the $\rm MgSiO_{3}$ condensation line crosses the blue T-P profile. 
}
\label{datafit}
\end{center}
\end{figure*}

Figure \ref{datafit} (right) shows the predictions of the blue T-P profile (Pont et al., 2013).  This profile produces a cloud top inflection point at about 0.3$\mu$m, and the steep slope at the shortest wavelengths is caused by hot $\rm H_{2}$ gas, but the enstatite cloud {\it base} is too deep in the atmosphere, giving $\lambda_D > 2 \mu$m and a transit that would obscure the K gas absorption line. 

Overall, the possible $0.6 \mu$m inflection may be consistent with an enstatite haze with abundance much smaller than cosmically available, for a T-P profile not too different from that of Fortney et al. 2008, but the steepness of the wavelength dependence at $\lambda <\lambda_D$, if interpreted as a temperature of 2000K, is not consistent with $\rm MgSiO_{3}$ as the responsible haze forming material.  The blue T-P profile has a thick enstatite cloud and a warm upper atmosphere, but is not consistent with a $0.6 \mu$m inflection and obscures the K line.

On a speculative note, we remark that the low abundance (relative to cosmic) we obtain for the tiny haze particles which cause the inflection at $\lambda_D \sim 0.6 \mu$m, {\it and} the moderately flat spectrum for $\lambda > \lambda_D$, might suggest a vertically extended haze of tiny (perhaps enstatite) particles overlying a highly settled enstatite cloud with large particles containing most of the Mg mass, lying very close to $z_c$ (like the settled clouds of AM01). The opacity of this settled layer might not be large, depending on the particle size, but could be wavelength independent, with a small effective scale height, and might create the moderately flat spectrum for $\lambda > \lambda_D$. As the data improves, perhaps these possibilities may be explored. 


\section{Summary}
We have extended the analyses of LeCavelier et al. (2008) to demonstrate the observable effects of condensation cloud bases on transit spectra. We showed how the signature of a condensation cloud base is an inflection or step at some $\lambda_{D}$ in the apparent radius spectrum. The inferred T-P value at the inflection provides important constraints on condensate identification. The magnitude and wavelength of the inflection can provide an estimate of the abundance of the condensate (however, this value depends on the poorly known haze particle size). The single cloud base development presented here is easily extended to multiple cloud bases. The spectrum of HD189733b could be viewed as having a cloud base inflection; more data are needed.

\end{document}